\documentclass[aps,amsmath,amssymb,twocolumn,superscriptaddress,citesort,longbibliography, pre, floatfix]{revtex4-2}
\usepackage{graphicx}% Include figure files
\usepackage{dcolumn}% Align table columns on decimal point
\usepackage{bm}% bold math
\usepackage{orcidlink}
\usepackage{xcolor}
\usepackage{hyperref}
\definecolor{tab_blue}{HTML}{1F77B4}
\hypersetup{
  breaklinks=true,
  colorlinks=true,
  linkcolor=tab_blue,
  filecolor=tab_blue,
  urlcolor=tab_blue,
  citecolor=tab_blue,
  pdfpagemode=FullScreen
}

\usepackage{float}
\floatstyle{plaintop}
\restylefloat{table}
\makeatletter
\let\newfloat\newfloat@ltx
\makeatother
\usepackage{algorithm}
\usepackage{algpseudocode}
\usepackage{setspace}
\usepackage{lipsum}
\usepackage{booktabs} 
\usepackage{tikz}
\usetikzlibrary{arrows.meta,calc}
\usepackage[percent]{overpic}

\definecolor{particle_color_1}{HTML}{488268}
\definecolor{particle_color_2}{HTML}{4D3D6E}
\definecolor{particle_color_3}{HTML}{B15B39}
\tikzset{
  particle_1/.style={
    circle,
    draw=particle_color_1,
    fill=particle_color_1,
    line width=0.35pt,
    inner sep=0pt,
    minimum size=0.32em
  },
  particle_2/.style={
    circle,
    draw=particle_color_2,
    fill=particle_color_2,
    line width=0.35pt,
    inner sep=0pt,
    minimum size=0.32em
  },
  particle_3/.style={
    circle,
    draw=particle_color_3,
    fill=particle_color_3,
    line width=0.35pt,
    inner sep=0pt,
    minimum size=0.32em
  },
  particle edge/.style={
    draw=black!65,
    line width=0.45pt,
    line cap=round
  },
  particle hidden edge/.style={
    draw=black!35,
    line width=0.35pt,
    dashed,
    line cap=round
  }
}
\newsavebox{\singlebox}
\newsavebox{\pairbox}
\newsavebox{\tetradbox}

\AtBeginDocument{%
  \sbox{\singlebox}{%
    \tikz[baseline=-0.45ex]{
      \node[particle_1] at (0,0) {};
    }%
  }%
  \sbox{\pairbox}{%
    \tikz[baseline=-0.45ex, x=1em, y=1em]{
      \draw[particle edge] (0,0) -- (0.85,0);
      \node[particle_2] at (0,0) {};
      \node[particle_2] at (0.85,0) {};
    }%
  }%
\sbox{\tetradbox}{%
  \tikz[baseline=-0.50ex, x=1em, y=1em]{
    \coordinate (A) at (0,0);
    \coordinate (B) at (1.15,0);
    \coordinate (C) at (0.575,0.996);
    \coordinate (D) at (0.575,0.38);
    \draw[particle edge] (A) -- (B);
    \draw[particle edge] (B) -- (C);
    \draw[particle edge] (C) -- (A);
    \draw[particle edge] (D) -- (A);
    \draw[particle edge] (D) -- (B);
    \draw[particle edge] (D) -- (C);

    \node[particle_3] at (A) {};
    \node[particle_3] at (B) {};
    \node[particle_3] at (C) {};
    \node[particle_3] at (D) {};
  }%
}
}
\DeclareRobustCommand{\single}{\usebox{\singlebox}}
\DeclareRobustCommand{\pair}{\usebox{\pairbox}}
\DeclareRobustCommand{\tetrad}{\usebox{\tetradbox}}

\begin{document}

% \preprint{APS/123-QED}

\title[Learning turbulent transport]{Learning turbulent transport via Mori--Zwanzig graph neural networks}

\author{Andr\'e Freitas\,\orcidlink{0009-0008-5256-7670}}
\email{andre.freitas@roma2.infn.it}
\affiliation{Department of Physics and INFN, University of Rome ``Tor Vergata'', Rome, Italy}
\affiliation{Information Processing and Communications Laboratory, T\'el\'ecom Paris, Institut Polytechnique de Paris, Palaiseau, France}

\author{Xander M. de Wit\,\orcidlink{0000-0002-7731-0598}}
\affiliation{Fluids and Flows Group and J.M. Burgers Center for Fluid Mechanics, Eindhoven University of Technology, Eindhoven, The Netherlands}

\author{Alessandro Gabbana\,\orcidlink{0000-0002-8367-6596}}
\affiliation{Department of Physics and Earth Sciences, University of Ferrara and INFN Ferrara, Ferrara, Italy}

\author{Michael Woodward\,\orcidlink{0000-0003-1481-3638}}
\affiliation{Computational Physics and Methods Group (CAI-2), Los Alamos National Laboratory, Los Alamos, NM, USA}

\author{Federico Toschi\,\orcidlink{0000-0001-5935-2332}}
\affiliation{Fluids and Flows Group and J.M. Burgers Center for Fluid Mechanics, Eindhoven University of Technology, Eindhoven, The Netherlands}
\affiliation{CNR-IAC, I-00185 Rome, Italy}

\author{Yen Ting Lin\,\orcidlink{0000-0001-6893-8423}}
\affiliation{Information Science Group (CAI-3), Los Alamos National Laboratory, Los Alamos, NM, USA}

\author{Daniel Livescu\,\orcidlink{0000-0003-2367-1547}}
\affiliation{Computational Physics and Methods Group (CAI-2), Los Alamos National Laboratory, Los Alamos, NM, USA}

\newcommand{\af}[1]{\textcolor{blue}{#1}}
\newcommand{\avg}[1]{\langle #1 \rangle}
\newcommand{\dd}{\mathrm{d}}
\newcommand{\vv}[1]{\bm{#1}}

\date{\today}

\begin{abstract}
We introduce a Mori--Zwanzig graph neural network (MZ--GNN) framework for learning reduced-order Lagrangian dynamics of tracer particles in homogeneous isotropic turbulence. The model represents particle acceleration as a finite-memory expansion over present and delayed particle-neighborhood graphs, with each memory contribution parameterized by an equivariant message-passing graph neural network. By construction, the architecture respects the relevant physical symmetries of the problem, including permutation equivariance, Galilean invariance, and equivariance under rotations and reflections. Trained on direct numerical simulation data, the model is rolled out autoregressively and evaluated on observables that are not imposed during training. We show that memory is essential for recovering the intermittent, heavy-tailed acceleration statistics, and that the learned dynamics accurately reproduce single-particle dispersion, pair-dispersion statistics, and four-particle tetrad geometry.  Our results establish a physically structured, scalable route to data-driven multi-particle simulation of turbulent transport, and a template for learning reduced dynamics of correlated, symmetry-rich particle systems.
\end{abstract}

\maketitle

%---------------------------------------------
\section*{Introduction}
%---------------------------------------------
Predicting the transport of material particles in turbulent flows remains a central challenge in fluid
dynamics~\cite{toschi-arfm-2009,coletti_rev}, with applications ranging from atmospheric pollutant and aerosol
dispersion~\cite{sawford-arfm-2001}, droplet growth in warm clouds~\cite{grabowski2013growth}, and fuel-air mixing in
combustion~\cite{pope1985pdf}, to ocean mixing and climate dynamics~\cite{lacasce2008statistics}.  
Despite decades of progress, a first-principles understanding of turbulent transport remains elusive because turbulence
is intrinsically multiscale, chaotic, and intermittent~\cite{frisch,alexakis2018,benzi_toschi_2023,sreeni25}: particle
trajectories are shaped by velocity fluctuations spanning many decades, from the integral length scale down to the
smallest dynamically active scales, the Kolmogorov length scales~\cite{falkovich_particles_2001, arneodo_universal_2008}, with
associated timescales ranging from the large-eddy turnover time to the fastest dissipative
times~\cite{pope-book-2001,yeung-arfm-2002,falkovich-nat-2002,toschi-arfm-2009}.

\begin{figure*}
  \centering
  \includegraphics[width=1.0\linewidth]{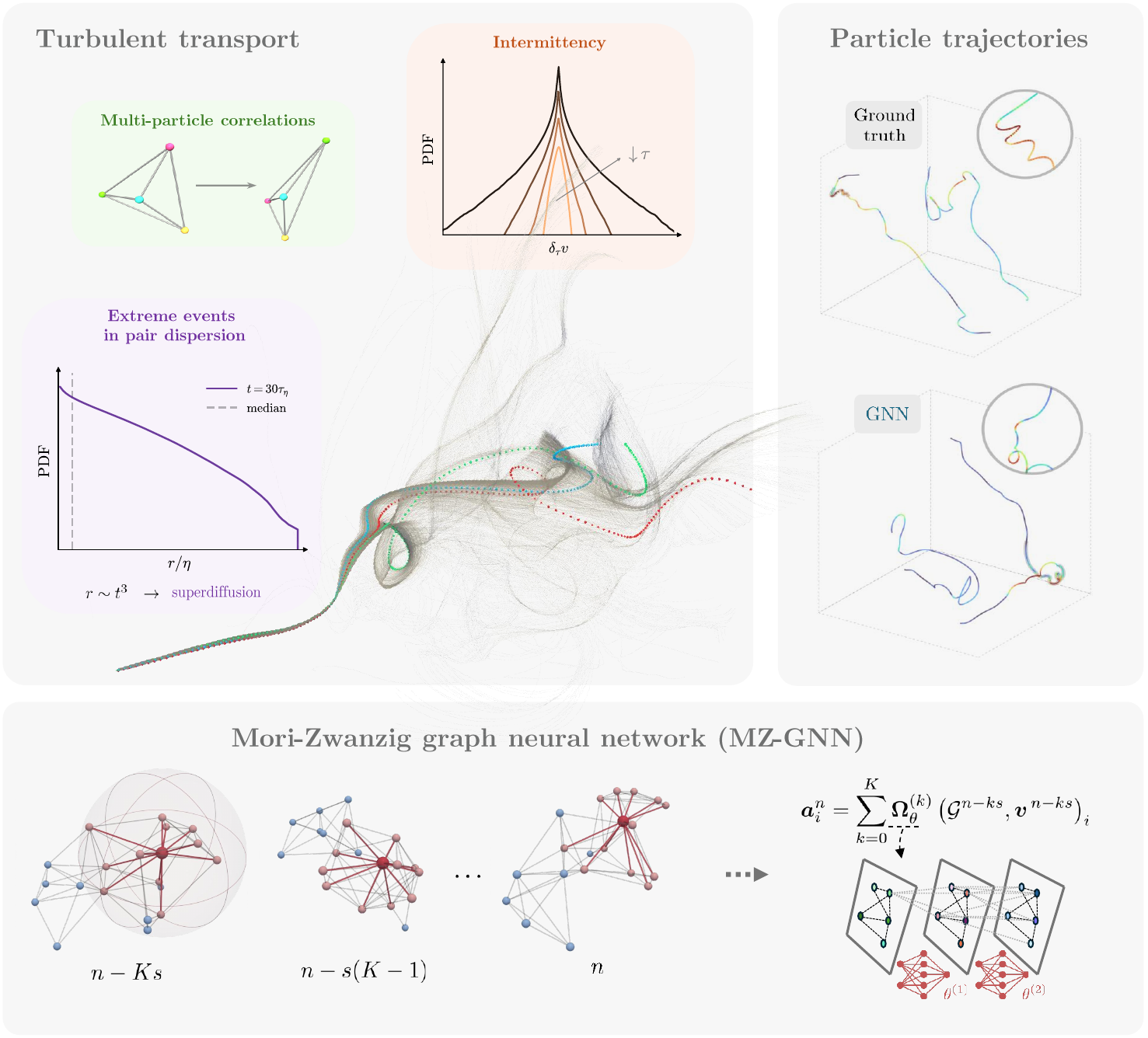}
    \caption{
    \textbf{Problem setup.}
    \emph{Turbulent transport.}
    A puff of passive tracers in homogeneous isotropic turbulence, dispersed throughout the turbulent flow across increasingly larger scales.
    The surrounding panels summarize key challenges for reduced-order Lagrangian modelling: intermittency, evidenced by the increasing non-Gaussianity of temporal velocity-increment probability density functions (PDFs) at decreasing time lags; rare extreme events in pair dispersion, where some particle pairs separate much faster than the median; superdiffusive separation; and non-trivial multi-particle correlations, illustrated by the deformation of a tetrad.
    \emph{Particle trajectories.}
    Representative Lagrangian trajectories coloured by acceleration magnitude comparing Eulerian--Lagrangian direct numerical simulation (DNS) ground truth with autoregressive graph neural network (GNN) rollouts, illustrating the complexity of tracer dynamics.
    \emph{Method.}
    Schematic of the Mori--Zwanzig graph neural network (MZ--GNN).
    Particles are represented as nodes of a graph, with edges encoding local particle neighborhoods defined by a cutoff radius.
    The highlighted red edges illustrate, for a given particle, the neighbours contained within the corresponding spherical interaction region.
    The particle acceleration is modelled through a finite-memory expansion over present and delayed graph states,
    $
    \vv{a}_i^n =
    \sum_{k=0}^{K}
    \boldsymbol{\Omega}^{(k)}_\theta
    \left(
    \mathcal{G}^{\,n-ks},
    \bm{v}^{\,n-ks}
    \right)_i
    $.
    The $k=0$ contribution is Markovian, whereas the $k>0$ terms encode non-Markovian memory effects.
    Each operator $\boldsymbol{\Omega}^{(k)}_\theta$ is parameterized by an equivariant message-passing graph neural network.
    }
  \label{fig:intro_panel}
\end{figure*}

The Lagrangian viewpoint, in which idealized massless particles are transported by the flow and tracked along their trajectories, provides a natural
framework for studying turbulent transport~\cite{biferale2006}.
Even at the single-particle level, turbulent dynamics exhibit striking signatures of intermittency. 
In particular, tracer accelerations display extreme fluctuations, with probability density functions characterized by heavy
non-Gaussian tails~\cite{vedula-pof-1999,laporta-nat-2001,voth-jfm-2002,sawford-pof-2003,mordant-physD-2004,biferale-pof-2005}.
These rare events reflect encounters with intense vortical and straining structures across the hierarchy of
turbulent scales~\cite{bec2006effects,bec-pof-2006,toschi-arfm-2009}. There have been many attempts to describe Lagrangian intermittency, with 
multifractal models arguably the most popular~\cite{borgas-prsa-1993,chevillard-prl-2003}. Other approaches include stochastic Langevin/Fokker-Plank models~\cite{pope-arfm-1994}, superstatistics~\cite{mordant-2004}, conditioning on coarse-grained dissipation or acceleration~\cite{bentkamp-natc-2019}, etc.
The phenomenology becomes richer in the multi-particle setting, where spatial correlations of the turbulent velocity field
play a central role. Pair dispersion quantifies how quickly turbulence separates initially close
particles~\cite{biferale2005lagrangian, bec2010turbulent} and provides a direct measure of turbulent mixing across
scales; four-particle clusters, or tetrads, additionally probe the deformation of material volumes by the local strain
field~\cite{chertkov1999lagrangian, pumir2000geometry, biferale2005multiparticle}. A faithful reduced model of
Lagrangian turbulence should therefore reproduce not only single-particle statistics but also pair and multi-particle
observables. Fig.~\ref{fig:intro_panel} illustrates the complexity of this transport problem. A puff
of tracer particles is released at very small separations and is subsequently stretched, folded, and dispersed across
scales by the turbulent flow, until its extent reaches the integral scale. The accompanying statistical diagnostics
highlight the range of observables that any successful reduced-order model must reproduce.

%===================================================================================================
%===================================================================================================

Direct numerical simulation (DNS), which solves the full Navier--Stokes equations and integrates tracer trajectories without
modelling assumptions, remains the reference approach for Lagrangian turbulence~\cite{riley-pof-1974,yeung-jfm-1989,squires-pofa-1991,yeung-jfm-2001,toschi-arfm-2009}. 
Its cost, however, grows as $N_{\mathrm{DOF}}\sim Re^{9/4}$ with the Reynolds number $Re=UL/\nu$ ($L$ a characteristic scale, $U$ a
characteristic velocity, and $\nu$ the kinematic viscosity), and quickly becomes prohibitive for the particle counts
demanded by realistic applications~\cite{yeung-pnas-2015,ishihara2016}. This has motivated decades of work on reduced-order descriptions
of particle transport. Classical probability density function methods~\cite{pope-arfm-1994,pope-book-2001,sawford-pofa-1991} represent turbulent transport
through ensembles of stochastic particles, governed by Langevin-type equations whose drift and diffusion terms
are modelled to reproduce selected turbulence statistics,
with extensions incorporating intermittency~\cite{wilson-blm-1996,lamorgese-jfm-2007,viggiano-jfm-2020}, synthetic velocity
fields from kinematic simulations~\cite{fung-jfm-1992,murray-pof-2016}, and cascade-inspired shell
models~\cite{biferale-arfm-2003}. Such models typically
rely on phenomenological assumptions and tunable closures, and they do not provide a general multi-particle dynamical
system for tracer transport. The growing availability of high-resolution DNS
databases~\cite{kaneda-pof-2003,yeung-pnas-2015,yeung-prf-2020} and experimental particle-tracking
data~\cite{schanz-ef-2016,schroeder-arfm-2023} has opened a complementary route: learning reduced-order models directly
from data. Machine learning, and deep learning in particular~\cite{Goodfellow-et-al-2016}, has achieved notable success in
Eulerian turbulence modelling, particularly in closure problems such as subgrid-scale modelling for large-eddy
simulation \cite{les_review,maulik2019subgrid,beck2019deep,ortali2024, freitas_PRF}. In the Lagrangian setting,
generative models~\cite{gilpin2024generative, carbone2024tailor, albergo_2025} such as diffusion model-based
approaches~\cite{GuastoniVinuesa2025} have recently produced synthetic single-particle trajectories with realistic
multiscale statistics, outperforming traditional stochastic models in statistical
fidelity~\cite{li-nmi-2024,li-ijmf-2024}. Such generative samplers, however, synthesize trajectories rather than
learning an explicit time-evolution law, which limits their use as autoregressive dynamical models for online coupling
with simulations, control, or prediction. While this strategy has been extended to particle
pairs~\cite{pantea2026turbulentpairdispersionstochastic}, scaling it to a complete multi-particle setting remains a
major challenge, since all spatial correlations must then be captured explicitly.

The definition of a reduced multi-particle surrogate for turbulent transport raises a closure problem. In DNS, tracer
trajectories are generated by the fully resolved Eulerian velocity field, whose state contains a very large number of degrees of freedom.
In contrast, the reduced model considered here has access only to the local information---such as positions, velocities, and accelerations---of a
finite number ($N_p$) of tracer particles, with $N_p$ much smaller than the number of degrees of freedom of the underlying
turbulent flow. These particle variables therefore do not define a closed Markovian state: the unresolved Eulerian
modes, together with flow structures and particles not represented in the reduced state, continue to influence the
future particle dynamics. 
The Mori--Zwanzig formalism~\cite{mori1965transport,zwanzig1973nonlinear} provides a principled description of
this situation: projecting the full dynamics onto the resolved observables yields an exact reduced equation containing a
Markovian term, a history-dependent memory term, and an orthogonal-dynamics term associated with the unresolved degrees
of freedom~\cite{chorin_optimal_2000,MZ_2010,zhen_md_2015,li2017computing,karthik_2017,gonzalez_learning_2020,lin2021datadriven_full,lin-siam-2023,Freitas_2026}. 
Importantly, the memory term and the orthogonal dynamics are not independent: they satisfy a self-consistent relation termed as the \textit{Generalized Fluctuation-Dissipation Relation} (GFD)~\cite{zwanzigNonequilibriumStatisticalMechanics2001,chorin_optimal_2000}.
Recent data-driven approaches have shown that the Markov and memory operators can be learned from time series of resolved
observables, with the learned memory constrained by the GFD~\cite{lin2021datadriven_full,lin-siam-2023}. This framework was recently applied to single-particle Lagrangian
turbulence~\cite{dewit-pnas-2026}, where the reduced dynamics of individual trajectories was learned using memory
operators augmented with time-delay embeddings. Because a single trajectory carries no information about its spatial
surroundings, that model had to incorporate the local flow structure explicitly: alongside the acceleration, it incorporated
velocity-gradient tensor at the particle's location as part of the reduced state (i.e., observables), as the velocity gradient is a proxy that encodes nearby strain and
vorticity that strongly shape tracer acceleration. This is an effective but indirect approach, reconstructing local spatial
information along a single trajectory at a time. Consequently, the model lacks the capability to represent and predict the joint spatial organization of many tracers whose
velocities and accelerations are correlated through the underlying velocity field.

A multi-particle surrogate must learn not only temporal memory but also how information is shared across particles in
physical space, a task for which geometric graph neural networks (GNNs) are naturally
suited~\cite{scarselli2009graph,gilmer2017neural,sanchez2020learning,bronstein2021geometric}. 
It has been shown that geometric GNNs, which build in physical symmetries such as permutation, translational, and rotational invariance and equivariance \cite{bronstein2021geometric}, can serve as powerful tools for learning complex interactions in physical systems such as molecular dynamics, granular media, and continuum mechanics~\cite{sanchez2020learning,pfaff2021learning}. Within the geometric GNN framework, we represent the particles as nodes of a graph embedded in continuous space, with local neighborhoods defining
edges and information exchanged through learned messages. The Mori--Zwanzig memory operators then act on present and
delayed particle-neighborhood graphs, so that temporal memory accounts for unresolved degrees of freedom while message
passing provides a learned, coarse representation of the local spatial structure of the flow. Spatial information is now
obtained directly from relative particle positions and velocities, hence the model no longer requires the
velocity-gradient tensor as an auxiliary input. To ensure that the learned representation is consistent with the underlying physics, we construct the graph encoding and message-passing architecture from relative particle states and symmetry-preserving geometric features. Because particle labels are arbitrary,  predictions must be i)
permutation equivariant; ii) since the flow is homogeneous, they must depend only on relative positions; iii) adding a
constant velocity is a change of inertial frame, hence they must be Galilean invariant and depend only on relative velocities;
and iv) because the flow is isotropic, vector-valued accelerations must transform equivariantly under rotations and reflections. By
building messages from relative positions, relative velocities, and scalar invariants, equivariant GNNs enforce all of
these constraints by construction~\cite{thomas2018tensor,fuchs2020se3,satorras2021en}.

%====================================================================================================
\begin{figure*}[!t]
  \centering
  \includegraphics[width=1.0\textwidth]{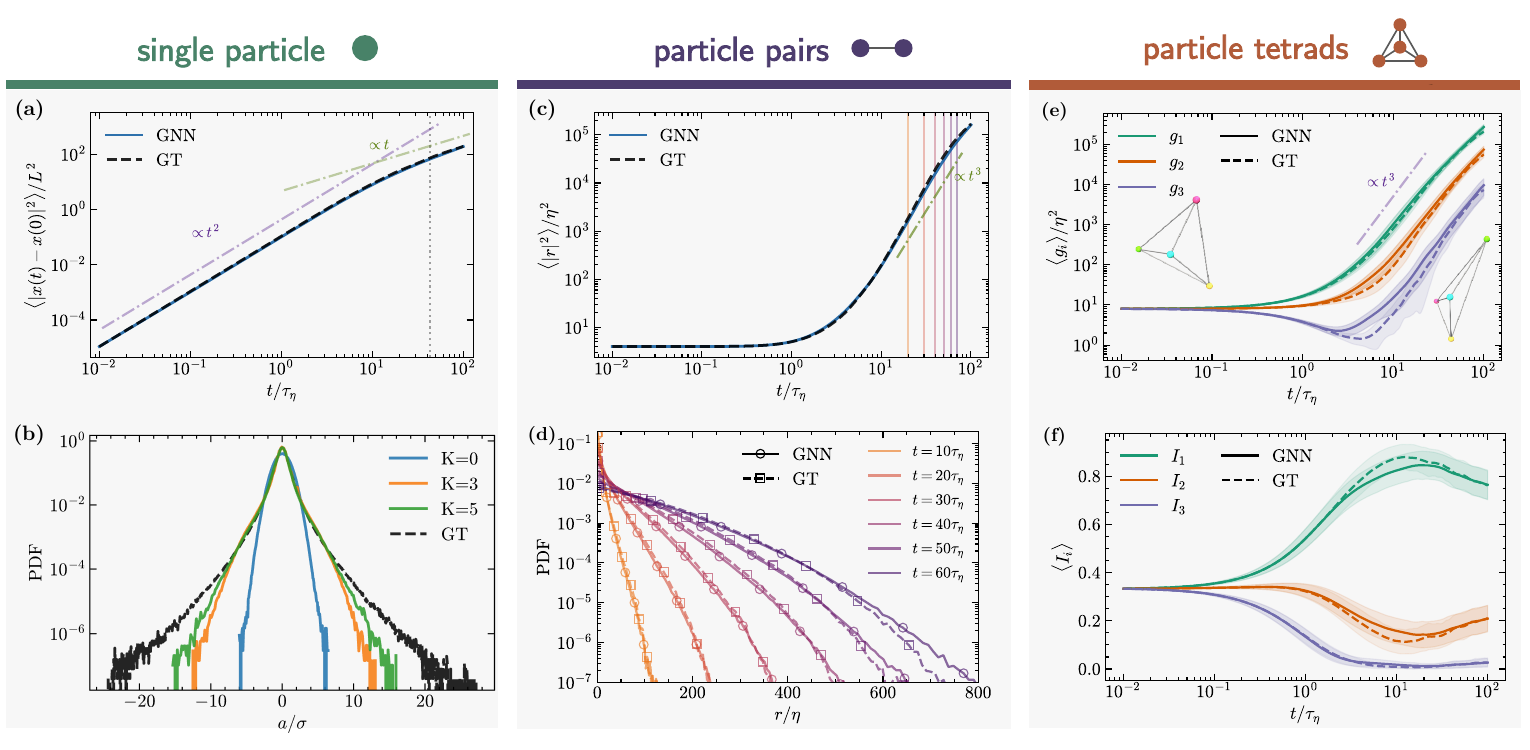}
\caption{
\textbf{Results}.
Single-particle, pair, and tetrad statistics from MZ--GNN rollouts compared with DNS ground truth.
(a) Mean-square single-particle displacement, showing the transition from ballistic scaling, $\sim t^2$, to long-time diffusive scaling, $\sim t$.
(b) PDF of the acceleration, normalized by its standard deviation, for memory depths $K=0$, $K=3$, and $K=5$.
Increasing the memory depth progressively partially recovers the non-Gaussian tails of the DNS.
(c) Mean-square pair separation, with a Richardson-like $\sim t^3$ superdiffusive scaling.
(d) PDFs of pair separation at different times, comparing MZ--GNN predictions and DNS.
(e) Mean eigenvalues $\langle g_i\rangle$ of the tetrad gyration tensor.
(f) Mean tetrad shape indices $\langle I_i\rangle$, with $I_i=g_i/(g_1+g_2+g_3)$.
Solid lines denote MZ--GNN predictions and dashed lines denote DNS ground truth unless otherwise indicated.
Unless explicitly varied, MZ--GNN results correspond to the full memory model with $K=5$.
In panels (e) and (f), shaded bands denote the estimated standard deviation.
}

  \label{fig:absolute}
\end{figure*}
%====================================================================================================

In this work, we introduce a Mori--Zwanzig graph neural network (MZ--GNN) for tracer dynamics in homogeneous isotropic
turbulence. The method extends data-driven MZ modeling from single-particle Lagrangian dynamics to a genuine
multi-particle setting. Its novelty is the combination of three ingredients: a finite-memory MZ expansion that closes
the dynamics of a sparse set of resolved particles; a graph-based representation that learns spatial correlations
directly from particle configurations; and exact physical symmetries enforced by construction. The acceleration of each particle is represented as a sum
of Markovian and memory contributions over present and delayed particle-neighborhood graphs. Each contribution is
parameterized by an equivariant message-passing GNN and trained sequentially in a residual fashion, following the
regression-based data-driven MZ paradigm developed in \cite{lin-siam-2023}. Representative DNS and model trajectories, together with a schematic of the
learning framework, are shown in Fig.~\ref{fig:intro_panel}. The model is then rolled out autoregressively and evaluated
on a broad set of Lagrangian observables: single-particle absolute dispersion, the heavy-tailed acceleration
distribution, pair-separation statistics, and four-particle tetrad geometry. 
The learned dynamics reproduces all of these observables, and we show that the non-Markovian terms are essential to
recover Lagrangian intermittency. Together, these results establish finite-memory, graph-based surrogates as a
physically structured and scalable route to data-driven multi-particle modeling of turbulent transport and a template
for learning reduced dynamics of correlated, symmetry-rich particle systems more broadly.

%---------------------------------------------
\section*{Results}
%---------------------------------------------

We assess the learned dynamics through rollouts of the trained graph neural network model.  Unless stated otherwise,
``GNN'' denotes the full model composed of one Markovian operator and five memory operators, i.e. $K=5$. A complete
detailed description of our approach is provided in the \textit{Methods}. To assess whether the model has learned the
true underlying turbulent dynamics, the evaluation is deliberately performed on observables that are not imposed as
losses during training: single-particle dispersion and acceleration statistics, pair-dispersion statistics, and the
geometry of four-particle clusters.  These observables probe increasingly stringent aspects of Lagrangian turbulence,
from one-particle temporal dynamics to multi-particle spatial correlations and deformation.  All rollouts at evaluation
time are performed in the periodic domain for a duration $T_{\mathrm{inf}}=100\tau_\eta$, where $\tau_\eta$ is the Kolmogorov time scale.  During the GNN evolution,
particle positions are wrapped periodically and graph edges are constructed using the minimum-image convention at each
time step based on a fixed cutoff radius.  For displacement-based observables, such as absolute dispersion, pair
separation, and tetrad growth, the corresponding unwrapped displacements are reconstructed by accumulating the
minimum-image increments along the trajectories. Unless otherwise specified, rollouts use $8000$ particles.  For PDF
estimates, where substantially larger statistical samples are required, we use ensembles containing up to $8\times 10^6$
particles.  Owing to memory constraints, these large ensembles are not evolved in a single GNN rollout but are split
into independent batches of particles, keeping the number of nodes fixed.  For single-particle statistics, particles are
initialized uniformly at random in the periodic cube.  For pair statistics, one particle is placed uniformly at random
in the domain and the second is initialized at a fixed separation $2\eta$ along the $x$ direction, with periodic
wrapping applied if necessary.  For tetrad statistics, a random centre is selected in the domain and four particles are
placed at alternating vertices of a cube, $(a,a,a)$, $(a,-a,-a)$, $(-a,a,-a)$, and $(-a,-a,a)$, relative to this centre.
With $a=\ell_0/\sqrt{2}$ and $\ell_0=2\eta$, this gives a regular tetrahedron with edge length
$\ell=2\sqrt{2}a=2\ell_0=4\eta$.

\paragraph*{Single-particle statistics. \single} 

We first consider one-particle observables, which test whether the learned dynamics reproduces the temporal statistics of individual tracer trajectories.
For a tracer initially at position $\vv{x}(0)$, the absolute displacement is
$
  \Delta \vv{x}(t)
  =
  \vv{x}(t)-\vv{x}(0),
$
and the corresponding mean-square displacement is
$
  D(t)
  =
  \left\langle
  |\Delta \vv{x}(t)|^2
  \right\rangle .
$
The largest contribution to the absolute tracer velocity comes from the velocity fluctuations at the integral scale. Hence, at short times and until the Lagrangian velocity has decorrelated, the displacement is approximately
$\Delta \vv{x}(t)\simeq \vv{v}(0)t$, leading to the ballistic regime
$
  D(t)
  \simeq
  \left\langle |\vv{v}|^2 \right\rangle t^2 .
$
Around the integral timescale, the motion becomes diffusive and
$
  D(t)
  \sim
  t.
$
Figure~\ref{fig:absolute}(a) shows that the GNN reproduces both regimes and the crossover between them, despite being trained only on short rollouts (spanning a fraction of $\tau_\eta$) with a mean-squared-error loss on the predicted acceleration. We then examine the acceleration statistics. Tracer acceleration directly probes the small-scale and intermittent structures encountered along particle trajectories. In Fig.~\ref{fig:absolute}(b), we show the PDF of the acceleration, normalized by its standard deviation, for different memory depths. A model without memory, $K=0$, strongly underestimates the probability of extreme acceleration events. Adding memory terms broadens the distribution and progressively partially restores the heavy non-Gaussian tails characteristic of Lagrangian intermittency~\cite{laporta-nat-2001,sawford-pof-2003,biferale-pof-2005}. For $K=5$, the predicted PDF is in close agreement with the DNS over several decades of probability. This shows that the non-Markovian terms in the Mori--Zwanzig expansion are essential for capturing the extreme events experienced by tracer particles. 

\paragraph*{Pair dispersion. \pair}

We next turn to pair statistics, which probe whether the model has learned the spatial correlations of the underlying velocity field.
For two tracers $i$ and $j$, we define the relative separation and relative velocity as
\begin{equation}
  \vv{r}_{ij}(t)=\vv{x}_i(t)-\vv{x}_j(t),
  \qquad
  \delta\vv{v}_{ij}(t)=\vv{v}_i(t)-\vv{v}_j(t),
\end{equation}
so that
\begin{equation}
  \frac{\dd}{\dd t}\vv{r}_{ij}(t)=\delta\vv{v}_{ij}(t).
\end{equation}
Unlike absolute dispersion, which is controlled by the temporal decorrelation of the single-particle velocity, relative dispersion is controlled by velocity differences at the instantaneous pair separation.
It therefore probes the spatial structure of turbulence and is sensitive to the intermittent nature of the cascade, with some pairs remaining close for long times while others undergo rapid separation events.

The classical description of turbulent relative dispersion goes back to Richardson~\cite{richardson1926}, who proposed that pair separation behaves as a non-Fickian diffusion process with a scale-dependent eddy diffusivity.
In isotropic turbulence, the radial separation PDF can then be modelled by a scale-dependent Fokker--Planck equation,
\begin{equation}
  \partial_t p(r,t)
  =
  \frac{1}{r^2}
  \partial_r
  \left[
    r^2 \mathcal{D}(r)\partial_r p(r,t)
  \right],
\end{equation}
where $\mathcal{D}(r)$ is the turbulent eddy diffusivity.
Richardson's empirical $4/3$ law, later connected to Kolmogorov--Obukhov dimensional arguments~\cite{kolmogorov1941local,obukhov1941spectral}, gives
\begin{equation}
  \mathcal{D}(r) \sim \varepsilon^{1/3} r^{4/3},
\end{equation}
with $\varepsilon$ the mean energy-dissipation rate.
Indeed, using the inertial-range estimates
$\delta v(r)\sim(\varepsilon r)^{1/3}$ and
$\tau(r)\sim r/\delta v(r)\sim\varepsilon^{-1/3}r^{2/3}$,
one obtains
$\mathcal{D}(r)\sim \tau(r)\delta v(r)^2\sim\varepsilon^{1/3}r^{4/3}$.
This leads to the Richardson--Obukhov law
\begin{equation}
  R^2(t)
  =
  \left\langle
  |\vv{r}_{ij}(t)|^2
  \right\rangle
  \sim
  g\,\varepsilon\,t^3,
\end{equation}
where $g$ is the Richardson constant.
The same phenomenology predicts a self-similar radial separation PDF of the form
\begin{equation}
  p(r,t)
  \sim
  \frac{r^2}{(\varepsilon^{1/3}t)^{9/2}}
  \exp\left[
    -C\frac{r^{2/3}}{\varepsilon^{1/3}t}
  \right],
\end{equation}
corresponding to explosive, superdiffusive separation.
The Richardson--Obukhov picture is based on a non-intermittent inertial-range scaling and therefore predicts a self-similar separation PDF.
In real turbulence this ideal self-similarity is modified primarily by intermittency: velocity increments fluctuate more strongly at smaller separation scales, producing rare pairs that separate much faster, or much slower, than typical pairs. These effects are most visible in the tails of the separation distribution~\cite{ott-jfm-2000,boffetta-prl-2002,ishihara-pof-2002,biferale2005lagrangian,scatamacchia2012extreme,biferale2014intermittency}.
Finite-Reynolds-number effects further complicate this picture by limiting the extent of the inertial range and introducing dissipative- and integral-scale cutoffs, as well as sensitivity to the initial separation.

Figure~\ref{fig:absolute}(c) shows the mean-square pair separation.
The GNN follows the DNS from the initial stage to the rapid superdiffusive regime, consistent with Richardson-like growth.
This agreement indicates that the learned dynamics captures the correlated relative motion of nearby particles, rather than only the one-particle statistics.

A more challenging test is provided by the full radial separation PDF in Fig.~\ref{fig:absolute}(d).
We show $p(r/\eta,t)$ at
$t=10\tau_\eta,20\tau_\eta,\ldots,60\tau_\eta$, all within the time range where the mean-square separation displays Richardson-like growth.
As time increases, the distributions broaden and shift towards larger separations, reflecting the explosive character of turbulent relative dispersion.
The GNN reproduces this evolution with high accuracy, including the rare-event tails associated with pairs that separate much faster than typical ones.
At the latest times, in particular around $t=60\tau_\eta$, the model slightly overestimates the far tails, indicating a mild excess of very rapidly separating pairs.
Overall, the agreement of the PDFs shows that the learned dynamics captures not only the average Richardson-like growth, but also much of the distribution of extreme dispersion events driven by intermittent, history-dependent relative motion.

To further assess the learned relative dynamics, we examine the distribution of pair relative velocities.
For a particle pair, we decompose the relative velocity into components parallel and perpendicular to the instantaneous separation vector,
\begin{equation}
  \hat{\vv{r}}_{ij}
  =
  \frac{\vv{r}_{ij}}{|\vv{r}_{ij}|},
  \qquad
  u_{\parallel,ij}
  =
  \delta\vv{v}_{ij}\cdot \hat{\vv{r}}_{ij}.
  \label{eq:relative_velocity_parallel}
\end{equation}
The perpendicular contribution is obtained by projecting out the longitudinal part,
\begin{equation}
  \delta\vv{v}_{\perp,ij}
  =
  \delta\vv{v}_{ij}
  -
  u_{\parallel,ij}\hat{\vv{r}}_{ij}.
  \label{eq:relative_velocity_perp}
\end{equation}
For the transverse PDF, we consider the signed components of $\delta\vv{v}_{\perp,ij}$ along two orthonormal directions perpendicular to $\hat{\vv{r}}_{ij}$ and pool both components.
The longitudinal component $u_{\parallel}$ controls the instantaneous growth of the pair separation, since
\begin{equation}
  \frac{\dd}{\dd t}|\vv{r}_{ij}|
  =
  u_{\parallel,ij}.
  \label{eq:pair_growth_longitudinal_velocity}
\end{equation}
It is therefore directly sensitive to stretching and compression events along the pair axis.
At early times, the pairs are initialized at small separations, so positive longitudinal relative velocities correspond to particles that rapidly move apart, whereas negative values correspond to pairs moving even closer together.
Because the initial separation is already near the dissipative scale, there is more room for explosive separation than for further approach, leading to a strongly skewed longitudinal PDF.
This asymmetry is related to the irreversibility of turbulent pair dispersion and the associated breaking of time-reversal symmetry.
As time increases and the pairs sample larger separations, the longitudinal distribution becomes progressively less skewed.
The transverse components instead describe the relative motion responsible for pair rotation and reorientation.
At early times, they probe small-scale turbulent velocity differences and therefore display strongly non-Gaussian tails.
As the pair separation grows, the transverse relative velocity samples increasingly larger scales, where fluctuations are less intermittent, and the PDF correspondingly becomes closer to Gaussian.

\begin{figure}[!t]
    \centering
    \includegraphics[width=1.0\linewidth]{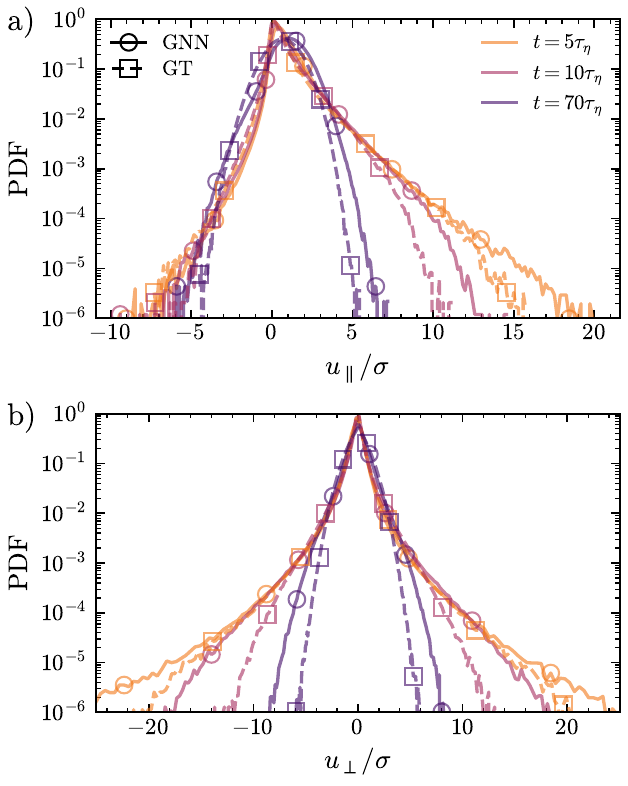}
    \caption{
    Pair relative-velocity statistics from MZ--GNN rollouts and DNS ground truth.
    (a) PDF of the longitudinal relative velocity $u_{\parallel}=\delta\vv{v}_{ij}\cdot\hat{\vv{r}}_{ij}$, normalized by its standard deviation.
    (b) PDF of the signed transverse relative-velocity components, obtained by projecting $\delta\vv{v}_{ij}$ onto the plane perpendicular to $\hat{\vv{r}}_{ij}$ and pooling the two transverse components, normalized by their standard deviation.
    Colours denote different times, $t=5\tau_\eta$, $10\tau_\eta$, and $70\tau_\eta$.
    Solid lines with circular markers denote MZ--GNN predictions, while dashed lines with square markers denote DNS ground truth.
    }
    \label{fig:pairs_vel}
\end{figure}

Figure~\ref{fig:pairs_vel} shows the PDFs of the longitudinal and transverse relative velocities at different times.
The MZ--GNN captures the overall evolution of both distributions, including the skewness of the longitudinal component and the progressive change in tail behaviour with time.
The agreement is particularly good for the transverse components, although the model slightly overestimates rare extreme events in both the longitudinal and transverse tails.
For the longitudinal component, this corresponds to a mild over-amplification of strong separation events, which is consistent across the times shown.
This local excess in positive relative velocity helps explain the overprediction of rapidly separating pairs observed at late times in the pair-separation PDFs of Fig.~\ref{fig:absolute}(d).
Thus, the relative-velocity statistics provide a more direct diagnostic of the same effect: the model learns the pair relative dynamics well overall, but slightly over-amplifies rare strong stretching and rotational events.

The effect of memory on a higher-order pair-dispersion statistic is reported in Appendix~\ref{app:memory_multiparticle}.

\paragraph*{Tetrad geometry. \tetrad}

Finally, we examine four-particle clusters, or tetrads.
Tetrads provide a higher-order test of the learned multi-particle dynamics because they probe the deformation of an extended material element, rather than only the separation of a pair.
While pair dispersion measures the growth of a single distance, tetrads also encode the shape of a material volume and therefore provide direct information on the geometry of turbulent deformation~\cite{chertkov1999lagrangian,pumir2000geometry,biferale2005multiparticle,hackl_PoF_2011,devenish2013geometrical}.
In turbulent flows, such clusters evolve both in size and in shape: their overall extent grows as particles separate, whereas the initially isotropic configuration is distorted by the local velocity-gradient field.

Given four particle positions $\{\vv{x}_a\}_{a=1}^4$, we define the tetrad centroid as
\begin{equation}
  \vv{x}_c
  =
  \frac{1}{4}
  \sum_{a=1}^{4}
  \vv{x}_a ,
\end{equation}
and the centroid-relative coordinates
\begin{equation}
  \bm{\rho}_a
  =
  \vv{x}_a-\vv{x}_c .
\end{equation}
The tetrad geometry is characterized by the gyration tensor
\begin{equation}
  \mathsf{G}
  =
  \frac{1}{4}
  \sum_{a=1}^{4}
  \bm{\rho}_a \bm{\rho}_a^{\top}.
\end{equation}
Let $g_1\geq g_2\geq g_3$ be its eigenvalues.
Their sum,
\begin{equation}
  r^2
  =
  \mathrm{Tr}\,\mathsf{G}
  =
  g_1+g_2+g_3,
\end{equation}
measures the overall tetrad size, while the normalized shape indices
\begin{equation}
  I_i
  =
  \frac{g_i}{g_1+g_2+g_3},
  \qquad
  \sum_{i=1}^3 I_i = 1,
\end{equation}
describe the tetrad shape independently of its size.
For a regular tetrad, the three eigenvalues are equal and $I_1=I_2=I_3=1/3$.
A flattened pancake-shaped configuration corresponds to $g_1\sim g_2 \gg g_3$, while an elongated needle-shaped configuration corresponds to $g_1\gg g_2,g_3$.
The shape indices therefore provide a compact diagnostic of the anisotropic stretching and compression induced by turbulence.

The physical picture is as follows.
At very small separations, the velocity field can be linearized around the tetrad centroid, so that the centroid-relative coordinates evolve approximately as
\begin{equation}
  \frac{\dd}{\dd t}\bm{\rho}_a
  \simeq
  \mathsf{A}\bm{\rho}_a,
  \qquad
  \mathsf{A}=\nabla\vv{u}(\vv{x}_c,t).
\end{equation}
This gives
\begin{equation}
  \frac{\dd}{\dd t}\mathsf{G}
  \simeq
  \mathsf{A}\mathsf{G}
  +
  \mathsf{G}\mathsf{A}^{\top}.
\end{equation}
The antisymmetric part of $\mathsf{A}$ mainly rotates the tetrad, while the symmetric strain-rate tensor controls stretching and compression. 
Because the flow is incompressible, the strain has zero trace, so extension along one direction must be accompanied by compression or weaker extension in the others. 
Initially regular tetrads are therefore rapidly distorted by the eigenstructure of the local strain field.
As the tetrad grows beyond the dissipative range, the relevant deformation is no longer controlled by the infinitesimal velocity gradient alone, but by a scale-dependent, coarse-grained strain over the size of the cluster. 
The shape dynamics then reflects a balance between coherent straining at the tetrad scale, which promotes anisotropic deformation, and incoherent relative motion from smaller scales, which tends to randomize the particle positions~\cite{chertkov1999lagrangian,pumir2000geometry}. 
Previous studies showed that inertial-range tetrads approach a statistically non-trivial shape distribution, with a preference for elongated and nearly coplanar configurations rather than uniformly distributed particle positions~\cite{biferale2005multiparticle,hackl_PoF_2011,devenish2013geometrical}. 
In analogy with Richardson pair dispersion, the overall tetrad size is expected to grow approximately as
\begin{equation}
  \left\langle \mathrm{Tr}\,\mathsf{G} \right\rangle
  \sim
  \varepsilon t^3,
\end{equation}
with the individual eigenvalues following the same growth up to shape-dependent prefactors in a self-similar regime.

The mean eigenvalues are shown in Fig.~\ref{fig:absolute}(e). 
Starting from regular tetrads, the DNS displays rapid growth of the cluster size together with a strong splitting of the three eigenvalues. 
The largest eigenvalue grows fastest, while the smallest remains much smaller, showing that the initially isotropic material element is stretched along a dominant principal direction and compressed or weakly stretched in the transverse plane. 
The GNN reproduces the DNS evolution of all three eigenvalues, indicating that it captures not only the growth rate of the cluster, but also the anisotropic deformation of its principal axes.
The shape indices in Fig.~\ref{fig:absolute}(f) isolate the geometry of the tetrad independently of its overall size. 
The increase of $\langle I_1\rangle$ and the simultaneous decrease of $\langle I_2\rangle$ and $\langle I_3\rangle$ show that initially regular tetrads evolve towards strongly distorted configurations. 
The GNN follows this anisotropization process and the subsequent time evolution with good accuracy. 
This agreement at the level of tetrad geometry provides strong evidence that the model has learned the strain-mediated, multi-particle deformation dynamics of the turbulent flow, rather than only matching scalar dispersion measures.

%================================================================================================
\section*{Discussion and conclusion}
%================================================================================================

In this work, we have introduced a physics-structured surrogate model for multi-particle Lagrangian turbulence, combining the Mori--Zwanzig formalism~\cite{mori1965transport,zwanzig1973nonlinear,lin-siam-2023,lin2021datadriven_full} with equivariant graph neural networks~\cite{gilmer2017neural,thomas2018tensor,fuchs2020se3,satorras2021en}.  The central idea is to represent tracer acceleration as a finite-memory expansion, where each memory contribution is parameterized by a message-passing GNN acting on present or delayed particle-neighborhood graphs, fully respecting the physical symmetries of the system.  This construction yields an explicit autoregressive dynamical system for tracer particles, rather than a purely generative model for trajectory statistics.
The key result of this reduced-order model is the accurate reproduction of multi-particle statistics.  While single-particle observables probe the temporal dynamics along individual trajectories, pair and tetrad statistics test whether the model has learned the spatially correlated structure of the turbulent velocity field.  The GNN reproduces the growth of pair separations, the evolution of the full pair-separation PDFs, and the rare-event tails associated with rapidly separating pairs.  The agreement extends to four-particle tetrads, where the model captures the growth of the gyration-tensor eigenvalues and the evolution of the normalized shape indices.  This is a demanding test: tetrad geometry depends not only on the overall rate of dispersion, but also on the anisotropic deformation of material elements by the turbulent strain field~\cite{chertkov1999lagrangian,pumir2000geometry,biferale2005multiparticle,hackl_PoF_2011,devenish2013geometrical}.  The fact that the model recovers these observables indicates that it has learned non-trivial multi-particle Lagrangian dynamics, beyond matching scalar one-particle statistics.

The results also highlight the importance of memory.  In the absence of memory, the model significantly underestimates the heavy tails of the acceleration distribution, failing to reproduce one of the defining signatures of Lagrangian intermittency~\cite{laporta-nat-2001,voth-jfm-2002,sawford-pof-2003,mordant-physD-2004,biferale-pof-2005}.  Adding Mori--Zwanzig memories progressively restores the non-Gaussian acceleration tails and leads to close agreement with DNS.  This supports the physical interpretation that unresolved turbulent degrees of freedom leave a measurable imprint on the resolved particle dynamics through history dependence. Consistent with recent data-driven Mori--Zwanzig approaches~\cite{lin-siam-2023,lin2021datadriven_full} and their application to single-particle statistics in turbulence~\cite{dewit-pnas-2026}, our results show that short-time supervised learning can yield stable long-time statistical predictions when the reduced dynamics are equipped with an appropriate memory structure. The symmetry properties of the architecture are equally important.  The model is permutation equivariant with respect to particle labels, Galilean invariant because it depends on relative velocities, and $O(3)$-equivariant because vector messages are built from equivariant bases with scalar coefficients depending only on invariant quantities.  These constraints encode the geometric and kinematic structure of homogeneous isotropic turbulence directly into the model class.  As a consequence, the network does not need to learn these symmetries from data, reducing the risk of orientation-dependent or frame-dependent artifacts and improving its ability to generalize to particle configurations not encountered during training.

More broadly, this work provides a route towards data-driven Lagrangian turbulence modelling in which the surrogate is both dynamical and multi-particle. 
Unlike purely generative approaches, the model advances the particle state step by step and can therefore be used in autonomous rollouts, coupled simulations~\cite{heart-turb}, or control-oriented settings~\cite{colabrese2017PRL, calascibetta2023taming}. 
Unlike single-particle closures, the graph formulation explicitly represents spatial correlations.
In this sense, the MZ--GNN can combine spatial and temporal information in a controlled way.
If all spatial degrees of freedom of the turbulent flow were resolved, the dynamics could in principle be represented by a Markovian evolution of the full state, as in DNS.
At the opposite extreme, a model based only on local single-particle information, such as Ref.~\cite{dewit-pnas-2026}, must rely on memory over extremely long times to account for correlations mediated by the unresolved flow.
The present approach bridges between these two extremes.
Spatial correlations that are resolved by the particle cloud are represented explicitly through graph interactions, with the largest accessible scales set by the graph receptive field.
Unresolved spatial information, including small-scale structures that would require a much denser particle sampling, is instead represented through a limited MZ memory.
This provides a natural division of labor: graph message passing captures the resolved spatial organization of the particles, while memory accounts for the unresolved degrees of freedom. The combination of memory, graph-based spatial interactions, and exact symmetries therefore appears to be a natural structure for reduced-order models of turbulent transport.

This perspective also connects the present framework to particle-based discretizations of continuum fluid dynamics.
Although the model is trained only on tracer trajectories and does not explicitly evolve an Eulerian velocity field, increasing the particle density would make the graph state a progressively richer representation of the underlying flow.
This places the approach close in spirit to smoothed particle hydrodynamics (SPH), where continuum fields are represented and evolved through particle degrees of freedom~\cite{Monaghan_2005,Monaghan_2012}, and suggests extensions towards graph-based Mori--Zwanzig surrogates or closures for particle methods, in the spirit of recent neural SPH approaches~\cite{toshev2024neural}.

The versatility of the approach presented here opens many new opportunities across a wide range of domains. The framework can be applied in many engineering settings where prediction or control of turbulent transport of passive tracers is of interest, yet where performing DNS is computationally prohibitive. It can also be extended to more complex forms of turbulent transport, such as inertial particles, where preferential concentration, caustics, and finite response times introduce additional complications~\cite{eaton1994preferential, bec2006acceleration, Calzavarini_prl2008, bec2010turbulent, dewit2025dynamicssmallbubblesturbulence}, making a robust data-driven approach such as that presented here even more valuable. Furthermore, coupling the learned Lagrangian surrogate with large-eddy simulations could provide a practical route to subgrid particle modelling~\cite{pozorski2009filtered, mazzitelli_PoF_2014,tian2023_lles}, where unresolved turbulent fluctuations must be represented without resolving the full Eulerian field.  Finally, the graph-based formulation opens the possibility of learning reduced dynamics for more complex particle systems, including active particles~\cite{ramaswamy2010mechanics,alert2022active}, droplets~\cite{girotto2024lagrangian}, or particles in inhomogeneous and anisotropic turbulent flows~\cite{biferale2016coherent, particles_tcf}.

%================================================================================================
\section*{Methods}
%================================================================================================

We describe the construction of the MZ--GNN model used to learn reduced-order tracer dynamics from DNS data.
We build on the regression-based data-driven Mori--Zwanzig framework developed in Refs.~\cite{lin2021datadriven_full,lin-siam-2023} and recently applied to single-particle Lagrangian turbulence in Ref.~\cite{dewit-pnas-2026}.
In this approach, the Markovian and memory operators of the Mori--Zwanzig expansion are learned from trajectory data, with regression defining an empirical projection onto the chosen resolved variables and function class.
The residuals left by this projection are associated with the orthogonal dynamics, and consistency of the learned memory terms is connected to the generalized fluctuation--dissipation relation, which links memory to correlations of the orthogonal dynamics.
Here, we extend this data-driven MZ construction to a multi-particle setting by representing each memory operator with a symmetry-preserving message-passing graph neural network acting on particle-neighborhood graphs.
The method therefore has three main components: first, the Mori--Zwanzig formalism motivates a finite-memory representation of the tracer acceleration; second, each memory operator is parameterized by an equivariant graph neural network; third, the memory operators are trained sequentially in a residual fashion and then rolled out autonomously to generate particle trajectories.
We first introduce the Mori--Zwanzig formalism and our data-driven formulation, then describe the graph neural architecture and its symmetry properties, and finally give the DNS and training details.

%================================================================================================
\subsection*{The Mori--Zwanzig framework}
%================================================================================================

The Mori--Zwanzig formalism~\cite{mori1965transport,zwanzig1973nonlinear} provides a systematic way of deriving closed, non-Markovian equations for a reduced set of observables of a high-dimensional dynamical system. Let $\bm{p}\in\mathbb{R}^{N}$ denote the full state and let $\bm{g}(\bm{p})\in\mathbb{R}^{M}$, with $M\ll N$, be the resolved observables. A projection operator $\mathcal{P}$ maps functions of the full state onto functions of the resolved variables alone. Applying the Mori--Zwanzig projection formalism yields the generalized Langevin equation (GLE)
\begin{equation}
  \frac{\dd}{\dd t}\bm{g}(t)
  =
  \bm{M}(\bm{g}(t))
  -
  \int_0^t
  \bm{K}(t-s,\bm{g}(s))\,\dd s
  +
  \bm{F}(t),
  \label{eq:GLE}
\end{equation}
where $\bm{M}$ is the Markovian contribution, $\bm{K}$ is the memory kernel, and $\bm{F}$ is the orthogonal-dynamics term. The latter represents the part of the full dynamics that is not determined instantaneously by the resolved variables. Equation~\eqref{eq:GLE} is exact for a given choice of observables and projection operator, but it is not directly usable as a reduced model because the memory kernel and the orthogonal dynamics are generally unknown.

In discrete time, the projected dynamics can be written as a non-Markovian autoregressive model,
\begin{equation}
  \bm{g}_{n+1}
  =
  \bm{\Omega}_0(\bm{g}_{n})
  +
  \sum_{k=1}^{n}
  \bm{\Omega}_k(\bm{g}_{n-k})
  +
  \bm{W}_n ,
  \label{eq:disc_GLE_full}
\end{equation}
where $\bm{\Omega}_0$ is the Markovian operator, $\bm{\Omega}_k$ are discrete memory operators, and $\bm{W}_n$ is the discrete orthogonal-dynamics contribution. The data-driven Mori--Zwanzig approximation consists of replacing the formally exact but unknown operators by learned regression functions and truncating the memory to a finite length $K$. Neglecting the explicit orthogonal-dynamics term then gives the closed truncated model
\begin{equation}
  \bm{g}_{n+1}
  \approx
  \bm{\Omega}_0(\bm{g}_{n})
  +
  \sum_{k=1}^{K}
  \bm{\Omega}_k(\bm{g}_{n-k}) .
  \label{eq:disc_GLE_truncated}
\end{equation}
This approximation should not be interpreted as assuming that the unresolved dynamics are absent.
Rather, their systematic influence on the resolved variables is represented through the learned memory operators.
In the MZ--GNN, the regression is performed over the class of equivariant graph functions, so that the empirical projection is restricted to functions of the resolved particle graph that satisfy the required physical symmetries.
The residual errors generated during learning play the role of the unresolved orthogonal dynamics, and the consistency of the learned memory terms is connected to the generalized fluctuation--dissipation relation~\cite{lin-siam-2023,lin2021datadriven_full}.

In the present work, the resolved variables are the positions and velocities of a set of tracer particles,
\begin{equation}
  \bm{g}_n =
  \left\{
  \left(\vv{x}_i^n,\vv{v}_i^n\right)
  \right\}_{i=1}^{N_p}.
\end{equation}
Rather than learning a direct map for the full next state, we parameterize the Mori--Zwanzig closure at the level of the particle acceleration. The learned acceleration is written as a finite-memory expansion over present and delayed particle states,
\begin{equation}
  \vv{a}_i^n
  =
  \sum_{k=0}^{K}
  \boldsymbol{\Omega}^{(k)}_{\theta}
  \left(
  \mathcal{G}^{\,n-ks},
  \bm{v}^{\,n-ks}
  \right)_i ,
  \label{eq:mz_acceleration}
\end{equation}
where $\mathcal{G}^{\,n-ks}$ is the particle graph constructed from the positions at the delayed time $n-ks$, $\bm{v}^{\,n-ks}$ denotes the corresponding velocities, $s$ is the memory stride, and $K$ is the maximum memory depth. The term $k=0$ is the Markovian contribution, while $k>0$ terms encode non-Markovian corrections arising from unresolved degrees of freedom.

Once the acceleration is predicted, the particle state is advanced explicitly via Euler as
\begin{subequations}
\label{eq:euler_update}
\begin{align}
  \vv{v}_i^{n+1}
  &=
  \vv{v}_i^{n}
  +
  \Delta t\,\vv{a}_i^n,
  \label{eq:euler_update_v}\\
  \vv{x}_i^{n+1}
  &=
  \vv{x}_i^{n}
  +
  \Delta t\,\vv{v}_i^n .
  \label{eq:euler_update_x}
\end{align}
\end{subequations}
with positions wrapped periodically when constructing the graph. The DNS acceleration provides the supervised target for training the learned closure.

Each operator $\boldsymbol{\Omega}^{(k)}_{\theta}$ in Eq.~\eqref{eq:mz_acceleration} is represented by a message-passing graph neural network. This is the main extension relative to single-particle data-driven Mori--Zwanzig models: instead of applying neural networks independently to individual trajectories, the memory operators act on a graph of particles and therefore encode spatial correlations between tracers. This construction allows the same framework to describe single-particle, pair, and multi-particle observables.

%=======================================================================================================
\subsection*{GNN architecture and symmetries}
%=======================================================================================================

Message-passing graph neural networks provide a natural representation for multi-particle Lagrangian dynamics. At each time, particles are represented as nodes of a graph and local interactions are encoded through edges. A message-passing layer updates the state of each node by aggregating information from its neighbours, using the same learnable function on every edge. This weight sharing makes the architecture permutation equivariant: relabeling the particles relabels the outputs in the same way. The learned dynamics should also respect the geometric and kinematic symmetries of the underlying Navier--Stokes equations. In homogeneous isotropic turbulence, the acceleration of a tracer particle is invariant under global translations, invariant under Galilean transformations, and equivariant under rotations and reflections of space. We therefore construct each learned operator so that it depends only on relative positions and relative velocities, and so that vector-valued messages are built from equivariant vector bases with scalar coefficients depending only on invariant quantities.

The model parameterizes the discrete GLE using a family of equivariant message-passing GNNs. 
For a given memory lag, each particle $i$ is a node. 
The graph edge displacement $\vv{r}_{ij}$ is defined as the shortest periodic displacement from particle $i$ to particle $j$, i.e. the minimum-image version of $\vv{x}_j-\vv{x}_i$.
Directed edges are introduced for $j\in\mathcal{N}(i)$ whenever $j\neq i$ and $\|\vv{r}_{ij}\|<r_c$, where $r_c$ is the graph cutoff radius and $\mathcal{N}(i)$ denotes the set of neighbours of particle $i$. 
We also define the Galilean-invariant edge velocity difference as $\vv{v}_{ij}=\vv{v}_j-\vv{v}_i$. 
The GNN takes the particle configuration and velocities as input and returns a vector contribution to the particle acceleration. 
The total acceleration is obtained by summing the contributions from the present and delayed graph states.

For a single memory operator, the latent vector feature of node $i$ at message-passing layer $\ell$ is denoted by $\vv{h}_i^{(\ell)}$. 
The latent vectors are initialized as
\begin{equation}
  \vv{h}_i^{(0)} = \mathbf{0},
\end{equation}
and updated over $N_{\mathrm{MP}}$ message-passing layers. 
At each layer, we form the relative latent vector
\begin{equation}
  \vv{h}_{ij}^{(\ell)}
  =
  \vv{h}_j^{(\ell)}-\vv{h}_i^{(\ell)} .
\end{equation}
The scalar edge features are chosen to be invariant under rotations and reflections:
\begin{align}
  s_{ij}^{(\ell)}
  =
  \bigl(
  &\|\vv{r}_{ij}\|^2,\;
  \langle \vv{r}_{ij},\vv{v}_{ij}\rangle,\;
  \|\vv{v}_{ij}\|^2,\; \nonumber\\
  &\langle \vv{r}_{ij},\vv{h}_{ij}^{(\ell)}\rangle,\;
  \langle \vv{v}_{ij},\vv{h}_{ij}^{(\ell)}\rangle,\;
  \|\vv{h}_{ij}^{(\ell)}\|^2
  \bigr).
  \label{eq:edge_invariants}
\end{align}
A shared multilayer perceptron (MLP)~\cite{Goodfellow-et-al-2016} $\phi_\ell$ maps these invariant scalars to scalar coefficients,
\begin{equation}
  \left(
  \alpha_{ij}^{(\ell)},
  \beta_{ij}^{(\ell)},
  \gamma_{ij}^{(\ell)}
  \right)
  =
  \phi_\ell\left(s_{ij}^{(\ell)}\right).
\end{equation}
The vector message from node $j$ to node $i$ is then constructed as a linear combination of equivariant vector bases,
\begin{equation}
  \vv{m}_{ij}^{(\ell)}
  =
  \alpha_{ij}^{(\ell)} \vv{r}_{ij}
  +
  \beta_{ij}^{(\ell)} \vv{v}_{ij}
  +
  \gamma_{ij}^{(\ell)} \vv{h}_{ij}^{(\ell)} .
  \label{eq:equivariant_message}
\end{equation}
Node features are updated by mean aggregation of incoming messages,
\begin{equation}
  \vv{h}_i^{(\ell+1)}
  =
  \vv{h}_i^{(\ell)}
  +
  \frac{1}{|\mathcal{N}(i)|}
  \sum_{j\in\mathcal{N}(i)}
  \vv{m}_{ij}^{(\ell)} .
  \label{eq:message_update}
\end{equation}
The acceleration contribution of this memory operator is read out from the final latent vector,
\begin{equation}
  \vv{a}_i^{(k)}
  =
  \vv{h}_i^{(N_{\mathrm{MP}})} .
  \label{eq:acc_readout}
\end{equation}
The spatial receptive field of each memory operator is controlled by the cutoff radius $r_c$ and the message-passing depth $N_{\mathrm{MP}}$.
The cutoff defines the one-hop neighborhood, while after $\ell$ message-passing layers node $i$ can depend on particles within at most $\ell$ graph hops.
Increasing $N_{\mathrm{MP}}$ therefore expands the receptive field at fixed cutoff, but very deep message passing can lead to oversmoothing or oversquashing of long-range information~\cite{rusch2023survey,arroyo2026vanishing}.
A more detailed discussion of the effective receptive field and the choice of $r_c$ and $N_{\mathrm{MP}}$ is given in Appendix~\ref{app:spatial_receptive_field}.

The symmetry properties follow directly from this construction. 
Since the same functions $\phi_\ell$ are applied on every edge and the neighbour aggregation is independent of the ordering of neighbours, the model is permutation equivariant. 
Since only relative positions enter, it is invariant under global translations. 
Since only relative velocities enter, it is invariant under Galilean transformations. 
Finally, for any orthogonal transformation $Q\in O(3)$, including rotations and reflections, the vectors
$\vv{r}_{ij}$, $\vv{v}_{ij}$, and $\vv{h}_{ij}^{(\ell)}$ transform as
\begin{equation}
  \vv{r}_{ij}\mapsto Q\vv{r}_{ij},
  \qquad
  \vv{v}_{ij}\mapsto Q\vv{v}_{ij},
  \qquad
  \vv{h}_{ij}^{(\ell)}\mapsto Q\vv{h}_{ij}^{(\ell)} ,
\end{equation}
while all scalar products and norms in $s_{ij}^{(\ell)}$ remain unchanged. 
Therefore the coefficients $\alpha_{ij}^{(\ell)}$, $\beta_{ij}^{(\ell)}$, and $\gamma_{ij}^{(\ell)}$ are invariant, whereas each message transforms as
$\vv{m}_{ij}^{(\ell)}\mapsto Q\vv{m}_{ij}^{(\ell)}$. 
By induction over message-passing layers, the acceleration contribution satisfies
\begin{equation}
  \vv{a}_i^{(k)}(Q\vv{x},Q\vv{v})
  =
  Q\vv{a}_i^{(k)}(\vv{x},\vv{v}) ,
\end{equation}
so each learned memory operator is $O(3)$-equivariant. 
The full Mori--Zwanzig model preserves the same symmetries, because the total acceleration is a sum of such equivariant GNN operators applied to present and delayed graph states.

%====================================================================
\subsection*{DNS data}
%====================================================================

The training and evaluation data are obtained from direct numerical simulations of incompressible homogeneous isotropic turbulence. 
The carrier flow evolves according to the Navier--Stokes equations
\begin{subequations}
\label{eq:NSE}
\begin{align}
  \partial_t \vv{u}
  +
  \vv{u}\cdot\nabla\vv{u}
  &=
  -\nabla p
  +
  \nu \nabla^2 \vv{u}
  +
  \vv{f},
  \\
  \nabla\cdot\vv{u}
  &=0,
\end{align}
\end{subequations}
in a triply periodic cubic domain of side $2\pi$. 
Here $\vv{u}(\vv{x},t)$ is the fluid velocity, $p$ is the pressure divided by the density, $\nu$ is the kinematic viscosity, and $\vv{f}$ is a large-scale forcing that maintains a statistically stationary turbulent state by imposing a constant energy-injection rate $\varepsilon$ on the low-wavenumber modes, with
$\hat{\vv{f}}(\vv{k})=\varepsilon\hat{\vv{u}}(\vv{k})/\sum_{k_f\leq|\vv{k}|<k_f+1}|\hat{\vv{u}}(\vv{k})|^2$ for $k_f\leq|\vv{k}|<k_f+1$~\cite{pope_forcing}.
The equations are solved with a standard pseudo-spectral method with $2/3$ dealiasing at resolution $N^3=256^3$, so that the maximum resolved wavenumber is $k_{\max}=N/3$. 
The simulation is fully resolved at the smallest scales, with $k_{\max}\eta \gtrsim 3$, and reaches a Taylor-scale Reynolds number $Re_\lambda\simeq 87$. A second-order accurate Adams--Bashforth scheme is used for time integration.

We consider passive tracers, corresponding to the zero-inertia, one-way-coupled limit of the Maxey--Riley--Gatignol equations~\cite{maxey_equation_1983, gatignol1983faxen}.
In this limit, particles are point-like and exactly follow the local fluid velocity,
\begin{subequations}
\label{eq:tracer_dynamics}
\begin{align}
  \frac{\dd \vv{x}_i}{\dd t}
  &=
  \vv{v}_i,
  \\
  \vv{v}_i(t)
  &=
  \vv{u}(\vv{x}_i(t),t).
\end{align}
\end{subequations}
The tracers do not exert feedback on the fluid and do not interact directly with one another.
The fluid velocity is interpolated to the particle positions using B-spline interpolation~\cite{michel_interp}.

The dataset contains particle positions, velocities, and accelerations for approximately $3.2\times 10^4$ tracers over $2.5\times 10^4$ stored time steps, corresponding to a total duration $T=250\tau_\eta$. 
The DNS integration time step is $\Delta t_{\mathrm{DNS}}=5\times 10^{-3}\tau_\eta$, while particle data are stored every two DNS steps. 
The effective time step used by the MZ--GNN model is therefore
$  
\Delta t = 10^{-2}\tau_\eta .
$
The dataset is split into training and testing subsets, using $80\%$ of the trajectories for training and $20\%$ for testing. 
During training, the stored trajectories are further arranged into short temporal windows, so that the model can be optimized through unrolled predictions over finite rollout horizons.

%====================================================================
\subsection*{Training details}
%====================================================================

The model is trained on short trajectory windows extracted from the DNS dataset. 
Each training sample consists of a time origin $n$, a batch of $N_p$ particles, and the delayed particle states required by the finite-memory expansion. 
At every time in the window, a periodic radius graph is built from the wrapped particle positions.
The graph cutoff is set to one third of the longitudinal integral length scale, $r_c=L_0/3$, with
$L_0=(3\pi/4)\int E(k)k^{-1}\,\dd k/\int E(k)\,\dd k$~\cite{pope-book-2001}, computed from the dataset.
The GNN therefore acts on periodic particle configurations, while unwrapped displacements are reconstructed only for diagnostics such as the ones shown in the results section.

For a given memory level $m$, the model receives the delayed graph states
$\{(\mathcal{G}^{n-ks},\bm{v}^{n-ks})\}_{k=0}^{m}$, where $s$ is the memory stride. 
The acceleration is predicted by the partial memory model containing the Markovian operator and the $m$ memory operators. 
This predicted acceleration is integrated forward with Eq.~\eqref{eq:euler_update} over a short rollout horizon $R$, rebuilding the graph after each position update. 
The loss is the mean-squared error between the predicted and DNS accelerations over the rollout,
\begin{equation}
  \mathcal{L}_m
  =
  \frac{1}{R N_p}
  \sum_{q=0}^{R-1}
  \sum_{i=1}^{N_p}
  \left\|
  \vv{a}_{i,m}^{\,n+q}
  -
  \vv{a}_{i,\mathrm{DNS}}^{\,n+q}
  \right\|^2 ,
  \label{eq:training_loss}
\end{equation}
where $\vv{a}_{i,m}^{\,n+q}$ denotes the acceleration predicted by the model at rollout step $q$. 

The memory operators are trained sequentially in a residual fashion. 
First, the Markovian operator $\boldsymbol{\Omega}^{(0)}_\theta$ is trained using only the current graph state. 
After convergence, it is frozen. 
For $m>0$, all previously trained operators
$\boldsymbol{\Omega}^{(0)}_\theta,\ldots,\boldsymbol{\Omega}^{(m-1)}_\theta$
are kept fixed, and only the new memory operator $\boldsymbol{\Omega}^{(m)}_\theta$ is optimized. 
Thus, each additional operator learns the residual acceleration not captured by the lower-order memory model. 
Optimization is performed with Adam~\cite{2015-kingma}.
The procedure is summarized in Algorithm~\ref{alg:mz_residual_training}, and the main hyperparameters are reported in Table~\ref{tab:training_hyperparameters}. 
After all memory operators are trained, the full model is rolled out autoregressively and evaluated on the single-particle, pair, and tetrad statistics reported above.

\begin{algorithm}[t]
\caption{Training of the MZ--GNN}
\label{alg:mz_residual_training}
\begin{algorithmic}[1]
\Require DNS trajectories $\{\vv{x}^n,\vv{v}^n,\vv{a}^n\}$, memory depth $K$, stride $s$, rollout horizon $R$
\For{$m=0,\ldots,K$}
  \State Initialise $\boldsymbol{\Omega}^{(m)}_{\theta}$
  \State Freeze $\boldsymbol{\Omega}^{(0)}_{\theta},\ldots,\boldsymbol{\Omega}^{(m-1)}_{\theta}$
  \For{each minibatch}
    \State Roll out the partial memory model up to level $m$ for $R$ steps
    \State Update only $\boldsymbol{\Omega}^{(m)}_{\theta}$ by minimizing $\mathcal{L}_m$
  \EndFor
\EndFor
\end{algorithmic}
\end{algorithm}

\begin{table}[t]
\centering
\caption{Main training and architecture hyperparameters.}
\label{tab:training_hyperparameters}
\begin{tabular}{lc}
\toprule
Quantity & Value \\
\midrule
Memory depth & $K=5$ \\
Memory stride & $s=5$ $\Delta t$ \\
Time step & $\Delta t=10^{-2}\tau_\eta$ \\
Graph cutoff & $r_c=L_0/3$ \\
Message-passing layers & $N_{\mathrm{MP}}=5$ \\
Hidden dimension & $256$ \\
MLP layers & $10$ \\
Total \# parameters & $\approx 10^7$ \\ 
Aggregation & Mean \\
Rollout horizon during training & $R=5$ \\
Learning rate & $5\times 10^{-4}$ \\
Particles per graph batch & $8000$ \\
\bottomrule
\end{tabular}
\end{table}
%==============================================================================================================
\section*{Data availability}
%==============================================================================================================
The data developed and used in this work are available from the corresponding author upon reasonable request.

\section*{Acknowledgements}
\raggedbottom
A.F. acknowledges useful discussions with Luca Biferale, Michele Buzzicotti and Maurizio Carbone. This research was supported in part by the European Union's HORIZON MSCA Doctoral Networks programme under Grant Agreement
No. 101072344, project AQTIVATE (Advanced computing, QuanTum algorIthms and data-driVen Approaches for science,
Technology and Engineering), the European Research Council (ERC) under the European Union's Horizon 2020 research and
innovation programme Smart-TURB (Grant Agreement No. 882340). This work is supported by the Netherlands Organization for
Scientific Research (NWO) through the use of supercomputer facilities (Snellius) under Grant No. 2025.008. This
publication is part of the project “Shaping turbulence with smart particles” with Project No. OCENW.GROOT.2019.031 of
the research program Open Competitie ENW XL which is (partly) financed by the Dutch Research Council (NWO). This work
has been co-authored by employees of Triad National Security, LLC which operates Los Alamos National Laboratory (LANL)
under Contract No. 89233218CNA000001 with the U.S. Department of Energy/National Nuclear Security Administration. We
acknowledge support from LANL's Laboratory Directed Research and development (LDRD) program, project number 20240318ER
and U.S.~Department of Energy, Office of Science, Office of Advanced Scientific Computing Research's Applied Mathematics
Competitive Portfolios program.

\section*{Author contributions}

A.F. developed the code, generated the training and testing datasets, and performed the numerical analysis.
All authors contributed to the conception of the research, interpretation of the results, and writing and revision of the manuscript.

\pagebreak

\bibliography{refs}% Produces the bibliography via BibTeX.

% \pagebreak
% \clearpage

\appendix

\section{Spatial receptive field}
\label{app:spatial_receptive_field}

The spatial information available to the MZ--GNN is controlled primarily by the graph cutoff radius $r_c$ and the number of message-passing layers $N_{\mathrm{MP}}$.
The cutoff radius defines the one-hop neighborhood, while $N_{\mathrm{MP}}$ determines how many graph hops information can travel before the acceleration is read out.
A simple geometric estimate of the maximum receptive field is
$
\ell_{\mathrm{eff}}
\sim
N_{\mathrm{MP}} r_c .
$
Writing $r_c=\alpha L_0$, with $L_0$ the longitudinal integral length scale, gives
$
\ell_{\mathrm{eff}}/L_0
\sim
\alpha N_{\mathrm{MP}} .
$
This should be interpreted only as an upper bound, since the actual information flow also depends on the instantaneous particle density and graph connectivity.

\begin{figure}[H]
\centering
\includegraphics[width=1.0\linewidth]{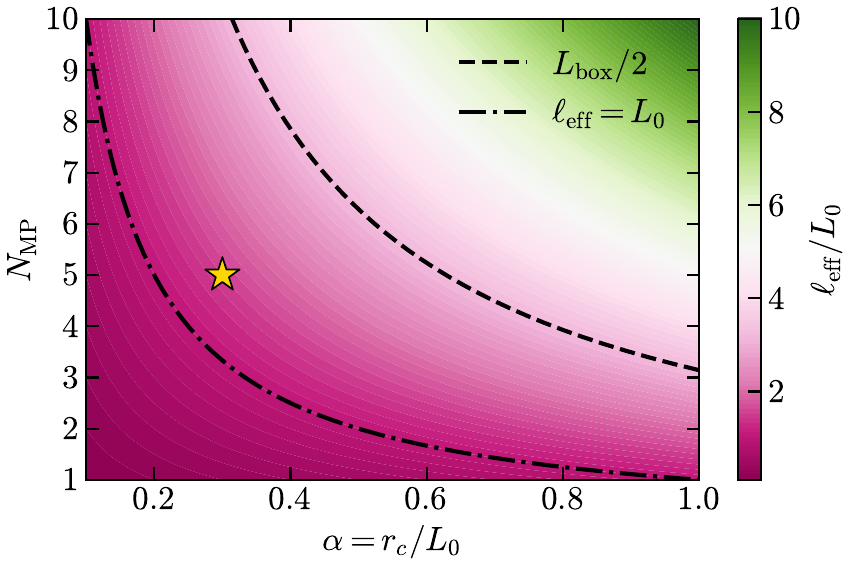}
\caption{
Effective spatial receptive field estimated as $\ell_{\mathrm{eff}}/L_0\sim\alpha N_{\mathrm{MP}}$, with $\alpha=r_c/L_0$.
The dash-dotted and dashed lines indicate $\ell_{\mathrm{eff}}=L_0$ and $\ell_{\mathrm{eff}}=L_{\mathrm{box}}/2$, respectively.
The star marks the configuration used in this work, $r_c=L_0/3$ and $N_{\mathrm{MP}}=5$.
}
\label{fig:spatial_receptive_field}
\end{figure}

Figure~\ref{fig:spatial_receptive_field} shows this estimate as a function of $\alpha=r_c/L_0$ and $N_{\mathrm{MP}}$.
The star marks the configuration used in this work, $r_c=L_0/3$ and $N_{\mathrm{MP}}=5$, corresponding to a receptive field of order the integral scale.
This choice balances two routes for increasing spatial information.
A larger cutoff gives direct access to a wider neighborhood, but increases graph density, computational cost, and may reduce locality, especially at high particle density.
Increasing the message-passing depth expands the receptive field at fixed local cutoff, but very deep GNNs can suffer from oversmoothing, where node features become less distinguishable, and from oversquashing, where long-range information is compressed into finite-dimensional node states~\cite{rusch2023survey,arroyo2026vanishing}.

The residual message-passing updates used here partially mitigate this issue, since each layer adds an incremental correction rather than replacing the node representation.
Higher-hop information therefore enters residually, supplementing shorter-range information rather than overwriting it.
Nevertheless, the choice of $r_c$, $N_{\mathrm{MP}}$, and particle density remains an important modelling decision: too small a receptive field limits the spatial correlations available to the model, whereas too large a graph may increase cost and degrade the learned local representation.
A systematic optimization over these parameters would require training many models and is left for future work.
Here, the chosen configuration should be understood as a physically motivated compromise between spatial coverage, locality, stability, and computational cost.

\section{Effect of memory on multi-particle statistics}
\label{app:memory_multiparticle}

As a complementary assessment of memory effects beyond the single-particle acceleration PDF discussed in the main text, we consider the flatness of the pair-separation distribution,
\begin{equation}
  F_r(t)
  =
  \frac{\left\langle |\vv{r}(t)|^4 \right\rangle}
       {\left\langle |\vv{r}(t)|^2 \right\rangle^2}.
\end{equation}
This quantity provides a measure of the tails of the separation PDF.
For a self-similar separation process, $F_r$ would remain approximately constant in time.
Changes in $F_r$ therefore indicate that the relative weight of typical and rare separation events is dynamical.

\begin{figure}[!t]
    \centering
    \includegraphics[width=1.0\linewidth]{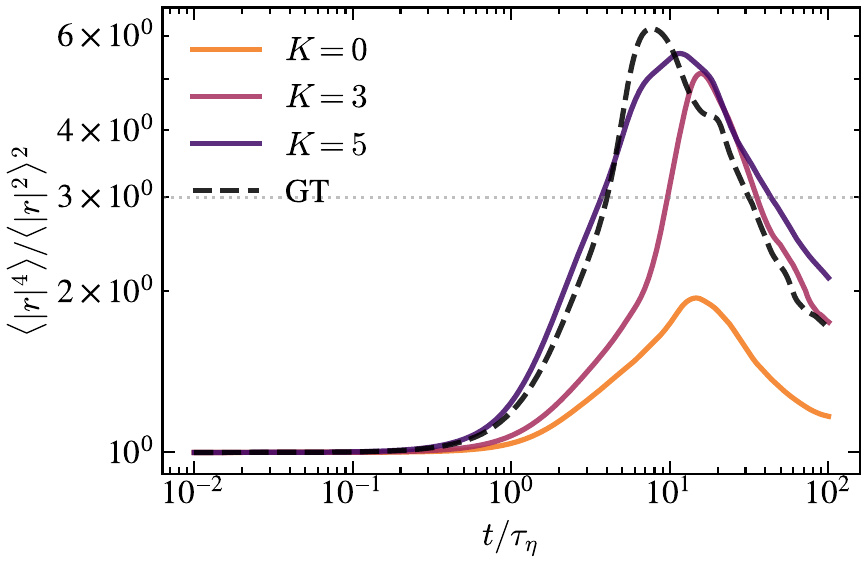}
    \caption{
    Flatness of the pair-separation distribution,
    $F_r=\langle |\vv{r}|^4\rangle/\langle |\vv{r}|^2\rangle^2$,
    for MZ--GNN models with different memory depths and DNS ground truth.
    }
    \label{fig:pairs_flatness}
\end{figure}

Figure~\ref{fig:pairs_flatness} shows $F_r(t)$ for different memory depths.
Starting from the initially fixed pair separation, the DNS flatness increases after the dissipative stage, reaches a maximum during the superdiffusive regime, and then decreases as pairs approach the large-scale diffusive regime.
At the present Reynolds number, these deviations from Gaussian-like behaviour are moderate, but they are still sensitive to the memory depth.
The Markovian model ($K=0$) strongly underestimates the growth of $F_r$, showing that memory is needed to capture the broadening of the pair-separation distribution.
Including memory substantially improves the agreement: both $K=3$ and $K=5$ reproduce the overall evolution, with $K=5$ giving the closest match to DNS.
The remaining differences are consistent with the slight overprediction of rapidly separating pairs discussed in the main text.
Overall, this confirms that memory improves not only single-particle statistics, but also higher-order multi-particle statistics of turbulent dispersion.

\end{document}